\documentstyle[12pt]{article}
\title{LATTICE CALCULATION OF GLUEBALL MATRIX ELEMENTS}
\author{ Y. Liang \thanks{Present address: Physics Dept., Univ. of
 Rochester, Rochester, NY 14627}\,\,, K.F. Liu, B.A. Li,
 and S.J. Dong \\ [0.5em]
 Dept. of Physics, Univ. of Kentucky, Lexington, KY
40506  \\[1em]
         K. Ishikawa \\ [0.5em]
         Dept. of Physics, Hokkaido Univ., Sapporo 060, Japan   }

\begin{document}

\maketitle

\begin{abstract}
Matrix elements of the form $ <0| Tr \; g^{2}GG |G> $
are calculated using the lattice QCD Monte Carlo method.  Here, $|G>$
is a glueball state with quantum numbers $ 0^{++}$, $ 2^{++}$, $
0^{-+}$ and $G$ is the gluon field strength operator. The matrix
elements are obtained from the hybrid correlation functions of the
fuzzy and plaquette operators performed on the $12^{4}$ and $14^{4}$
lattices at $\beta = 5.9 $ and $5.96$ respectively.  These matrix
elements are compared with those from the QCD sum rules and the
tensor meson dominance model. They are the non-perturbative matrix
elements needed in the calculation of the partial widths of
$J/\Psi$ radiative decays into glueballs.
\end{abstract}

%\section{INTRODUCTION}\label{introduction}

The glueballs predicted from Quantum Chromodynamics (QCD) are the most
exotic particles from the point of view of the naive quark model, but
they are crucial to the QCD picture of the hadronic interactions.  The
identification of certain experimentally observed particles as
glueballs will be a strong confirmation of QCD.  Radiative $J/ \Psi$
decays have been regarded as the best hunting ground for
glueballs\cite{kopke}.  Several candidates from experiments have
existed for some time\cite{wisniewski}.  In order to compare the
experimental findings with the theory, we need to find at least some
characteristics of the glueball spectrum, or better yet to calculate
some properties of the glueballs from QCD.   Previous theoretical
studies have concentrated on the estimation of the masses of the
glueballs.  The prediction of these glueball masses has long been
attempted in lattice gauge Monte Carlo calculations \cite{glu1}.
Recent progress using fuzzy loops \cite{Tep1,mt,dg87}, smeared loops
\cite{APE}, and inverse
 Dirac operators \cite{Sch1,kronfeld} to improve the signal to noise
ratio have yielded quite consistent results on large volumes. These
calculations show that the lowest scalar, tensor, and pseudoscalar
glueballs lie in the mass range of 1 - 2.5 GeV. In this letter,
we shall report
a lattice gauge Monte Carlo calculation of vacuum to glueball
transition  matrix elements which are needed for estimating the
partial widths of $J/ \Psi$ radiative decays into glueballs.

The glueball matrix elements we
compute are of the form $ <0| g^{2}G_{\mu\nu}^{a}G_{\rho\sigma}^{a}
(x) |G> , $ where $G$ can be the $ 2^{ ++ } $, $ 0^{ ++ } $ or $ 0^{
-+ } $ glueball and
 $G_{\mu\nu}$ is the gluon field strength operator.
 On a lattice, the local operator
$g^{2}G_{\mu\nu}^{a}G_{\rho\sigma}^{a} (x)$ can be constructed by
small loops consisting of lattice links.  One way to do the
calculation might be, say for the $2^{++}$ glueball, to calculate the
correlation function $< O(t) O(0) >$, where $ O = 2 \Box_{12} -
\Box_{31}- \Box_{23}$.  Here, $\Box_{\mu\nu}$ is a plaquette with
edges in the $\mu$ and $\nu$ directions.  Then, for large Euclidean
time,
\begin{equation}
< O(t) O(0) > \sim
 (< 0| O |1 >)^{2} e^{- m_{1}t} ,
 \end{equation}
 where $ |1> $ is the lowest glueball state with the quantum numbers
associated with O. The matrix element can thus be extracted.
Technically, however, we know from glueball mass calculations
\cite{Tep1}$^{-}$\cite{baal} that small loops usually are not very
good for creating glueball states.  The states they create have small
projections to the glueball states resulting in large noises in Monte
Carlo calculations.  In fact, it has become increasingly standard
\cite{Tep1,mt,dg87,baal} in mass calculations to use the
so-called ``fuzzy
operators'', which have extended structures, presumably reflecting the
size of the glueball.  On the other hand, it is difficult,
if at all possible, to relate
the fuzzy operators to the field strength operators numerically.
%-----------------------------------------------------------------------
%Here is the \raggedbottom \pagebreak
%keeps the second column from being written over
%-----------------------------------------------------------------------
Therefore, we adopt a method which is a hybrid of the two
approaches.  We use fuzzy operators to do a variational calculation
first, finding the linear combination of different fuzzy operators
that maximizes the overlap with a glueball state in the calculation of
the correlation function of the fuzzy operators $< O(t) O(0) >$.
Then, we calculate the correlation functions of small loops with the
fuzzy operator $O$
\begin{equation}
< O(t) FF(0) > =\! \sum_{n}\! <\! 0| O |n\! > <\!n| FF
 |0\!> e^{- m_{n}t} ,
 \end{equation}
  Here, $FF$ is the lattice approximation of the $ G^a G^a $ operator.
If we assume that the variational calculation gives a good overlap
with the lowest glueball state as we let $t >> a$, the lattice spacing,
then we could fit the two correlations to the following expressions
\begin{eqnarray}
 < O(t) O(0) > \sim <G|O |0>^2  e^{ -mt},  \label{eq1}  \\
 < O(t) FF(0) > \sim <G| O |0> <0| FF | G> e^{ -mt} \label{eq2},
\end{eqnarray}
for different time t with three parameters, i.e. $<G|O|0>, <0|FF|G>,$
and m. This way we could obtain the desired matrix element $<0|FF|G>$
and the glueball mass m. Since these two correlation functions are
measured on the same set of gauge configuraitons, we shall use the
data-covariance matrix to fit them simulataneously \cite{glr89,bhl92}.
More details to follow.

%mall loops, we expect that the correlation $ < O(t) FF(0) > $ can go
%less far in $t$ than $ < O(t) O(0) > $  before the error becomes too
%large.  Hence, in both of the above formulae, the time separation
%in the $<O$ $O >$ correlation is devised to be
%larger than that in the $< O$ $ FF >$
%correlation.  We will check to see if these two different definitions
%yield consistent results.

Gluon configurations are generated by a pseudo-heat-bath algorithm
\cite{cabibbo}.  A measurement is made for every ten configurations
generated.  For each configuration to be measured, fuzzy links are
constructed {\it a la} Michael \& Teper\cite{mt}, as in Fig.1.
%\begin{figure} \vspace{2truein} \caption{The making of fuzzy links
%{\it a la} Michael and Teper .} \end{figure}
The procedure produces
links with length twice of the original links.  It can be iterated
several times to yield links of length $ 2^{\ell}$.  In our
calculations, $\ell=2$.  After this procedure is done, the resultant
links are no longer $SU(3)$ matrices.  They do have the correct gauge
transformation properties, however.  In principle, these links can be
used to form loop operators to create glueball states.  In practice,
however, we have found that they lead to huge statistical
fluctuations.  We use the following steps to convert the links back to
$SU(3)$ matrices.  Two conditions should be satisfied.  First, if the
fuzzy link happens to be an $SU(3)$ matrix, the procedure should keep
the matrix un-altered.  Second, the final matrix should have the
correct gauge transformation properties.  Let us consider a matrix
$M^{\prime}$ and $ \det M^{\prime} \neq 0$.  We first divide it by its
determinant $ M= M^{\prime} / \det M^{\prime} $.  This obviously
satisfy the two conditions.  Now, any complex non-degenerate matrix
can be factorized uniquely into a positive definite Hermitian part on
the right and a unitary matrix on the left, $M= UH$(one can factorize
the
matrix as $M=H^{\prime} U^{\prime}$ too resulting in a different set
operators to create the glueball states).  If under a gauge
transformation $M \mapsto g_1 M g_{2}^{\dagger}$, by the uniqueness
for the factorization, $ U \mapsto g_1 U g_{2}^{\dagger}$ and
$ H \mapsto
g_2 H g_{2}^{\dagger}$.  Uniqueness also means that $H=1$ and $M=U$ if
$M$ is unitary.  Therefore, when we convert $M^{\prime}$ to $U$, both
conditions are satisfied.  In actual computation, we first find the
positive definite Hermitian matrix $H^{2} =  M^{\dagger} M$.  We use
the Jacobi method to diagonalize $H^{2}$ leading to
$ H^{2} T = T \Lambda$,
where $T$ is the orthonormalized unitary matrix of eigenvectors of
$H^2$ and $ \Lambda $ is the diagonal matrix of eigenvalues.
The inverse of H can be obtained from $H^{-1}
= T \Lambda^{-1/2} T^{\dagger}$.  Finally, $U= MH^{-1}$.  This
procedure is somewhat different from that used in \cite{Tep1,mt}.

%----------------------------------------------------------------------%
% Here is the \flushbottom %            to keep the bottoms of the
%columns flush
%----------------------------------------------------------------------%
To form the $O$ operators for various spins, we use the $U$ fuzzy
links to form loops with shapes depicted in Fig.2.  The same shapes
were used in \cite{mt}.
%\begin{figure} \vspace{3truein}
%\caption{Shapes of different loops.}
%\end{figure}
Using the method
outlined in \cite{johnson}, these loops and their rotations about
different axes thereof are
combined into operators belonging to the $A^{++}_1 (0^{++})$, $E^{++}
(2^{++})$, $T^{++}_2 (2^{++})$, $A^{-+}_1 (0^{-+})$ representations of
the cubic group.  The notations in the parentheses are the presumed
quantum numbers for $J^{PC}$
when the rotation invariance is restored.  Only the real part of
the trace is used because all the operators we consider here have
positive charge conjugation.  $(a)$ and $(b)$ in Fig. 2
contribute to all four kinds
of representations.  $(c)$ does not contribute to the pseudoscalar.
$(d) $ and $(e)$  only contribute to $A^{++}_1 (0^{++})$ and  $E^{++}
(2^{++})$.  These operators are averaged over spatial coordinates for
each time slice to obtain the zero-momentum projections.

The $G^a G^a $ operators are constructed from the elementary(not
fuzzy) links in Fig.3.  Because of all the different shapes of the
fuzzy operators and the yet-unknown variational coefficients, we do
not know the polarization of the tensor states created by the fuzzy
operators.  We therefore need to calculate all the components of the
$FF$ tensor.  For the pseudoscalar, we also need different components
of the tensor in order for the $F \tilde{F}$ pseudoscalar to have all
the required symmetry properties.  To get these, we first construct a
lattice approximation of the field strength tensor.
%\begin{figure}
%\vspace{2truein}
%\caption{Operators for local field strength operator.}
%\end{figure}
\begin{equation}
L_{\mu\nu} = L_1 + L_2 +L_3 +L_4 .
\end{equation}
where $L_i (i = 1,4)$ are products of link variables $UUU^+U^+$
around the plaquettes in the $\mu\nu$ plane as indicated in Fig. 3.
\begin{eqnarray}
F_{\mu\nu} &=&( L_{\mu\nu} -
L_{\mu\nu}^{\dagger} ) /8 \nonumber  \\
           &=& a^2 A_{\mu\nu} + {1 \over 6 } a^4 ( D_{\mu}^2 +
D_{\nu}^2 )  A_{\mu\nu} + O(a^6 ).
\end{eqnarray}
Here, repeated indices are
{\it not } summed over and $A_{\mu\nu} = i g G_{\mu\nu} $.  $a$ is the
lattice spacing.  Therefore
\begin{equation}
Tr ( F_{\mu\nu}
F_{\rho\lambda} ) \propto a^4 g^2 G_{\mu\nu}^{a} G_{\rho\lambda}^{a} +
O(a^6 ) .
\end{equation}
This way, we can find all the components of
the field strength tensor and all the Lorentz components of (gauge
 invariant) trace of products of the field strength tensor.  The fuzzy
operators are chosen according to the size and scale of the glueball.
The operators for the field strength tensor, on the other hand, have
sizes on the order of one lattice spacing.  Once we choose the lattice
scale such that the glueball sizes are much larger than the lattice
spacing, then we expect that terms in higher powers of the lattice
spacing $a$ are relatively unimportant.  This is the basis of our
strategy to use large, fuzzy links to create glueballs and measure the
field strength using small, elementary links. We shall use two $\beta$
values to check the scaling of the results.

A variational calculation which maximizes
\begin{equation}
{ c_i c_j < O_i (a) O_j (0) > \over  c_i c_j < O_i
(0) O_j (0) > } = {  < \Phi | e^{ - a H}  | \Phi > \over
  < \Phi | \Phi >  }
 \end{equation}
 is carried out to find the coefficients $\{ c_i \}$.
Here, $ | \Phi > = \sum_{i} c_{i} O_{i} |0> $ and $O_{i}$ are operators
corresponding to different loops in Fig. 2.  From the right hand
side of the equation, we see that this is really not different from
the usual variational calculation in quantum mechanics.  Instead of
taking the expectation value of the Hamiltonian itself and minimizing
it, we take the expectation value of $ e^{ -a H}$ and maximize it,
which corresponds to finding the lowest eigenvalue of $H$.  For the
scalar case, vacuum expectation values are first subtracted from the
correlations, which corresponds to making the variational state $|
\Phi >$ orthogonal to the vacuum state.  In actual calculation, this
is equivalent to finding the eigenvector for the largest eigenvalue
for the generalized eigenvalue problem,
\begin{equation}
  < O_i (a) O_j (0) > c_j =   \lambda  < O_i (0) O_j (0) > c_j .
\end{equation}
Once this is done, we  proceed to evaluate the
correlations $ < O(t) FF(0) > $ and
\mbox{$ < O(t) O(0) > $} using $ O= c_i
O_i$.  Both the variational calculation and the fitting of
Eqs. (\ref{eq1},\ref{eq2}) are used to extract the desired matrix
elements and glueball masses.
%\begin{equation} { < O(t) FF(0) > \over < O(2t) O(0) > ^{1/2} }
%\sim <0| FF | 1>, %\label{formula} %\end{equation}

For different spins, only the operators with same quantum numbers have
non-zero matrix elements between the vacuum and the glueball states.
  Specifically, we would like to extract the following matrix elements
for different quantum numbers
\begin{description}
\item[Scalar]
   \begin{eqnarray} \label{eq:scalar}
   s/a^{4}\! &=&\!<0| Tr (g^{2}G_{\rho\sigma}G_{\rho\sigma} )
   |G>    \nonumber \\
    &=& \! 2 <0| Tr g^{2}( E \cdot E + B \cdot B ) |G> .
\end{eqnarray}
\item[Pseudoscalar]
   \begin{eqnarray} \label{eq:pseudoscalar}
   p/a^{4}\!&=&\!  <0| \epsilon^{\mu\nu\rho\sigma} Tr
( g^{2}G_{\mu\nu} G_{\rho\sigma} ) |G> \nonumber \\
&=&\! 4 <0| Tr (g^{2}E \cdot B ) |G> .
\end{eqnarray}
\item[Tensor]
\begin{equation} \label{eq:tensor}
t \epsilon_{\mu\nu} / a^4 =  <0| \Theta_{\mu\nu} |G>
 \end{equation}
 where
$\epsilon_{\mu\nu}= \epsilon_{\nu\mu}$, $\epsilon_{\mu\mu}= 0$,
$\epsilon_{\mu\nu} \epsilon_{\mu\nu} =1$, and $ \Theta_{\mu\nu} = 2 Tr
g^2 ( G_{\mu\sigma} G_{\nu\sigma} - 1/4 \delta_{\mu \nu}
G_{\rho\sigma} G_{\rho\sigma} ) = g^2 T_{\mu \nu}$ where $T_{\mu\nu}$
 is the  energy-momentum tensor.
Here,
\begin{eqnarray}
\lefteqn{t^2/4\,g^4a^8 = ( < E_i E_j > - < B_i B_j
> )^2} \nonumber \\
 & & + 2\! <E_i B_j > (<E_i B_j > -  <E_j B_i > ) .
 \end{eqnarray}
 and $ < \ast > \equiv <0| \ast  |G>$.
\end{description}
The s, p, and t in the above equations are the lattice matrix elements
obtained from eqs. (\ref{eq1}) and (\ref{eq2}).

Up to now,  we have assumed the normalization
\mbox{$ <G|G> =1 $.} To go to the continuum
normalization \mbox{$ <G, {\bf p} |
G, {\bf p}^{\prime} > = 2 E_{\bf p} (2 \pi )^{3} \delta^{3} ( {\bf  p
- p^{\prime}} )$,}  \\
or \mbox{$ <G,0|G,0> = 2 M (2 \pi )^{3} \delta^{3}
( 0) = 2 M V $,} we need to multiply the above matrix elements by $
\sqrt{ 2 M V}$.

 Calculations are performed for a $ 12^4$ and a $14^4$ lattices with
$\beta$ value equals to $5.9$ and $5.96$ respectively. The lattice
sizes are chosen so that the L\"{u}scher scale parameter
$z = m_{A^{++}_1} L$
\cite{lusch} is about 10 where it is found \cite{baal} that $E^{++}$
and $T^{++}_2$ start to become degenerate indicating
the restoration of the rotational invariance. Furthermore, their
physical sizes are about the same. From a cold start, we
begin to make measurements after 4000 thermalization sweeps.  The
measurements are made for every 10 updates.  For the $ 12^4$ lattice
at $\beta= 5.9$, 5000 measurements have been performed.  For the $
14^4$ lattice at $\beta= 5.96$, 5800 measurements have been performed.
Errors that enter the data-covariance matrix for the $\chi^2$ fit of
eqs. (\ref{eq1}) and (\ref{eq2})
are estimated by binning the data into 20 sets and
doing jackknife on them.  The parameters chosen are in the scaling(not
necessarily asymptotic scaling) regime\cite{mt}.  We present the
results in units of the string tension $K$, which is determined from
fitting the correlations of Polyakov loops to the exponential
$ A e^{-KLt}$ for the time interval  t = 2 to 4 for $\beta = 5.9$
and t = 2 to 5 for $\beta = 5.96$.
 The scale determined from the string
tension indicates that the lattice spacing decreases by $\sim 19 \%$
from $\beta = 5.9$ to 5.96. This makes the physical sizes of the
$12^4$ and $14^4$ lattices about the same.
  The glueball masses determined from both the joint fit of the
correlations of eqs. (\ref{eq1}) and (\ref{eq2}) and the fit of the
correlation of fuzzy loops alone in eq. (\ref{eq1})
are listed in Table 1. They are given in units of $\sqrt{K}$.
They agree with each other and
with the higher statistics calculations in this $\beta$ range
\cite{mt} (i.e. $\beta = 5.9, 6.0$) for each
 mass determined from the corresponding time slices.
However, for the $T_2^{++}(2^{++})$ case, we could not find a fit
with small $\chi^2$. This presumably reflects the fact that our
lattice is still not large enough for the tensor glueball, since it is
known that the tensor glueball is about 4 times larger than the
scalar glueball~\cite{fl92,fl92a}.

%\vskip .5cm \leftline{Masses in unit of $\sqrt{K}$:}
\begin{table}
\caption[lst_entry]{Glueball Masses in unit of $\sqrt{K}$. $t_1 /
t_2$ denotes the range of time slices of eq. (\ref{eq1}) /
eq. (\ref{eq2}) that are fitted. Also included are the $\chi^2$
per degree of freedom}
\begin{tabular}{|c|c|c|l|c||c|c|l|c|}
\hline
 &   $t_1$ & $t_2$  &  $\beta\!=\!5.9 $  & $\chi^2 /N_{DF}$
 & $t_1$ & $t_2$ & $\beta\!=\!5.96 $ & $\chi^2 /N_{DF}$ \\
 \hline
 $A_{1}^{++} $
    & 4,5 & 1,2 & 3.08(1) & 0.1 &1,2,3 & 2 & 3.42(1) & 7.9 \\
$(0^{++})$
    & 1,2,3,4,5  &  & 3.29(9) & 0.14 & 1,2,3 &  & 3.40(3) & 0.04 \\
 \hline
 $E^{++} $
 &2,3  & 0,1 & 4.44(1) & 19.2 & 3.4 & 0,1 & 4.14(1) & 0.3 \\
$(2^{++})   $
  & 1,2,3 &  & 4.68(3) & 52.0 & 2,3,4 &  & 4.33(4) & 2.8 \\
\hline
$T_{2}^{++} $
  & 1,2 & 1,2 & 4.83(1) & 190 & 1,2,3 & 1 & 5.63(1) & 138 \\
$(2^{++})$
   & 1,2,3 &   & 5.12(2) & 242 & 1,2,3 &  & 5.72(18) & 72 \\
\hline
$A_{1}^{-+} $
   & 1,2,3 & 1  & 6.10(3) & 49.7 & 1,2,3 & 1 & 6.18(2) & 1.0 \\
$(0^{-+})$
   & 1,2,3 &  & 6.05(3)& 40  & 1,2,3 &  & 6.17(2) & 0.8 \\
\hline
\end{tabular}
\end{table}
\vskip .3cm
 The glueball matrix elements with several fits to eqs. (\ref{eq1})
and (\ref{eq2})
 are list in Tables 2 through 5. They are in the unit of $K^{3/2}$.

\begin{table}
\caption[lst_entry]{Scalar matrix elements obtained from
Eqs. \ref{eq1} and \ref{eq2}. They are in the unit of $K^{3/2}$.}
\begin{tabular}{|r|r|l|c||r|r|l|c|}
\hline
    \multicolumn{4}{|c||}{ $\beta=5.9$}
   & \multicolumn{4}{c|}{ $\beta=5.96$}\\
 \hline
 $t_1$ & $t_2$ & $ s/a^4$ & $\chi^2 / N_{DF}$ & $t_1$ & $t_2$ & $s/a^4$
 & $\chi^2 /N_{DF}$  \\
 \hline
 4,5 & 2,3  & 45.2(3) & 4.7 & 1,2,3  & 3& 51.1(4) & 7.94  \\
 4,5 & 1,2 &  47.2(3) & 0.1 & 1,2,3  & 2& 53.0(1)  & 7.9   \\
\hline
\end{tabular}
\end{table}

\begin{table}
\caption[lst_entry]{Tensor matrix elements for the $E^{++}$
representation in units of $K^{3/2}$.}
\begin{tabular}{|c|c|l|c||c|c|l|c|}
\hline
    \multicolumn{4}{|c||}{ $\beta=5.9$}
   & \multicolumn{4}{c|}{ $\beta=5.96$}\\
\hline
 $t_1$ & $t_2$& $ t/a^4$ & $\chi^2 / N_{DF}$ & $t_1$ & $t_2$ & $t/a^4$
 & $\chi^2 /N_{DF}$  \\
\hline
2,3 & 0,1  & 7.42(6) & 19.2 & 3,4 & 0,1 & 11.0(1)  & 0.3    \\
    &      &         &      & 2,3,4 & 2 & 8.64(14) & 4.5 \\
\hline
\end{tabular}
\end{table}

\begin{table}
\caption[lst_entry]{Tensor matrix elements for the $T_2^{++}$
representation in units of $K^{3/2}$.}
\begin{tabular}{|r|r|l|c||r|r|l|c|}
\hline
    \multicolumn{4}{|c||}{ $\beta=5.9$}
   & \multicolumn{4}{c|}{ $\beta=5.96$}\\
\hline
 $t_1$ & $t_2$& $ t/a^4$ & $\chi^2 / N_{DF}$ & $t_1$ & $t_2$ & $t/a^4$
 & $\chi^2 /N_{DF}$  \\
\hline
 1,2 & 1,2  & 6.38(8) & 190  & 1,2,3 & 1 & 10.1(1) & 138     \\
\hline
\end{tabular}
\end{table}

\begin{table}
\caption[lst_entry]{Pseudoscalar matrix elements in units of
$K^{3/2}$.}
\begin{tabular}{|c|c|l|c||c|c|l|c|}
\hline
    \multicolumn{4}{|c||}{ $\beta=5.9$}
   & \multicolumn{4}{c|}{ $\beta=5.96$}\\
\hline
 $t_1$ & $t_2$& $ p/a^4$ & $\chi^2 / N_{DF}$ & $t_1$ & $t_2$ & $p/a^4$
 & $\chi^2 /N_{DF}$  \\
\hline
 1,2,3 & 1 & 19.2(2) & 50  & 2,3 & 1,2 & 21.9(2) & 210 \\
 2,3   & 1,2 & 17.4(2) & 0.05 & 1,2,3 & 1 & 23.2(2) & 1.0 \\
\hline
\end{tabular}
\end{table}
These numbers can be converted into physical units using $ K \approx (
0.42 GeV )^2 $({\it i.e.}, multiplying the numbers by $K^{3/2}$).  We
see from these results that the scalar glueball matrix elements are
better calculated, much like the calculation of the glueball masses.
The variation from $\beta = 5.9$ to $5.96$ is about $10\%$. This is
quite reasonable as the matrix element scales like $K^{3/2}$. The
rest of the matrix elements differ more. Again, the case for
$T_2^{++}$ is the worst with no fit of small $\chi^2$, enough though
the values of the matrix element are not that dissimilar to that of
$E^{++}$. Both $E^{++}$ and $A_1^{-+} (0^{-+})$ have fits with
reasonably small $\chi^2$.

To come to grips with the continuum matrix elements,
we need to consider the finite
lattice renormalization of the lattice operators. The finite
renormalization of the lattice operators considered here have been
calculated perturbatively \cite{dg90,cdp88,cmp91,cfr90}. We can
define the renormalization of the operators as follows,
\begin{equation}
O = Z O_L a^{-d}
\end{equation}
where the $O_L$ is the lattice-regularized version of the continuum
operator $O$ and d is the mass dimension of
$O$. For the case of the scalar operator $g^2 G_{\mu\nu}
G_{\mu\nu}$ which coincides with the renormalization group invariant
trace anomaly
$\frac{\beta(g^2)}{g} G_{\mu\nu} G_{\mu\nu}$ to lowest order
in $g$, the renormalization constant Z is found \cite{dg90} to be
1.06 at one-loop order. For the pseudoscalar lattice operator
which is the topological charge in the limit of zero lattice spacing,
the finite renormalization Z is found to be 2.5 with the two-loop
corrections estimated by dominant tadpole graphs \cite{cdp88}.
For the energy-momentum tensor operator, the lattice perturbation
to one loop gives the Z to be 1.84 for $\beta = 5.9$ \cite{cfr90}.

When these renormalization constants are taken into account, the
scalar matrix element in eq.(\ref{eq:scalar}) is about
$50 K^{3/2}$
or about $3.7 GeV^3$ for $K = (0.42 GeV)^2$.
Based on the scaling properties of QCD and trace anomaly, both
the QCD sum rule \cite{nsvz80} and the soft meson
theorem \cite{el85} lead to the estimate that relates the
scalar glueball matrix element to the gluon condensate,
\begin{equation}
<0| Tr (g^{2}G_{\mu\nu}G_{\mu\nu} ) | G> =
16 \pi^2 \sqrt{\frac{G_0}{2b}} m_G ,
\end{equation}
where $G_0 = <0| \frac{\alpha_s}{\pi} G_{\mu\nu}^a G_{\mu\nu}^a |0>$
is the gluon condensate, $b = 11/3 N_c - 2/3 N_f$ and $m_G$ the
scalar glueball mass. Taking $N_f = 0$ for the pure gauge case,
$G_0 = 0.012 GeV^4$ from the ITEP value \cite{svz79}, and $m_G$
in the range of 1 to 1.7 GeV, this matrix element is then
estimated to be 3.7 $\sim$ 6.3 GeV. This agrees with our lattice
calculation to within a factor of two for the extreme cases.
For the pseudoscalar case, the lattice renormalization $Z_p = 2.5$
makes the glueball matrix element in eq. (\ref{eq:pseudoscalar})
to be about $44 \sim 58 K^{3/2}$ or $3.3 \sim 4.3 \, GeV^3$.
It has been proposed that there is an approximate chiral symmetry
between the scalar and pseudoscalr glueballs \cite{cs84}.
A sum rule is derived from an effective action which relates
the topological susceptibility $\chi$ in the pure gauge case
to the gluon condensate $G_0$
\begin{equation} \label{eq:sumrule}
\chi = \xi^{-2} \frac{G_0}{2b}
\end{equation}
where $\xi$ denotes the degree of chiral symmetry. From the
Witten-Veneziano \cite{wv79} mass formula for the $\eta'$ mass
which is related to $\chi$ in the large $N_c$ analysis,
this sum rule predicts $\xi$ to be 0.7. From our lattice results,
$\xi$ can be obtained from the ratio of the pseudoscalar to
scalar matrix elements \cite{cs84}, i.e.
\begin{equation}
\xi_L = \frac{ Z_p p/a^4}{2 Z_s s/a^4}.
\end{equation}
When we take the ratio between the pseudoscalar and scalar matrix
elements from Tables (2) and (5), $\xi_L$ truns out to range from
0.5 to 0.6. This is quite close to the prediction from the sum rule
in eq. (\ref{eq:sumrule}). In the Witten-Veneziano mass formula for
the $\eta'$ mass \cite{wv79}, the glueball contribution to the
topological susceptibility is neglected. The glueball contribution
to the topological susceptibility is
\begin{equation}
\chi_{glueball} = (\frac{<0|Tr(g^2 G \tilde{G})|G>}{16 \pi^2
 m_G} )^2
\end{equation}
Using our lattice result and the mass of the pseudoscalar glueball
at 1.4 GeV, the pseudoscalar glueball turns out to give
an appreciable 9 \% contribution to the topological susceptibility.
With the perturbative lattice renormalization $Z_T$, our lattice
result for the tensor glueball matrix element is
$\sim 14 K^{3/2}$ or $\sim 1.0 \, GeV^3$. On the other hand, the
prediction from the tensor dominance model \cite{itll88}
and QCD sum rule \cite{nar84}
is about $ 0.16 \sim 0.35 \, GeV^3$ for the tensor glueball mass
ranging from 1.7 to 2.2 GeV. Hence, the lattice result is about
3 to 6 times larger. Since the tensor dominance model gives an
reasonable prediction of $\theta(f_2(1720))$ production rate in
$J/\Psi$ radiative decay \cite{itll88,lsl87},
we suspect that the lattice result we obtain for the tensor
case is overestimated. Recent glueball wavefunction study
\cite{fl92,fl92a} reveals that the size of the tensor glueball is
about 4 times
larger than those of the scalar glueball and the pion \cite{lli89}.
Hence, we
expect that  the tensor glueball is more susceptible to the
finite size effect. Considering that our choice of the lattice
size should be large enough for the scalar and pseudoscalar
glueballs but may not be enough for the tensor, the
squeezed tensor glueball will lead to an overestimated tensor
matrix element. In the constituent glue picture, the matrix
element we calculate can be viewed
as the wavefunction at the origin which scales as the inverse
3/2 power of the glueball size. We think this is the likely
source for the large tensor matrix element.

Finally, we would like to discuss the prediction of the glueball
production rates in $J/\Psi$ radiative decays. While the details of the
calculation will be presented elsewhere \cite{lli_future},
we would like to mention some of the main features of the findings.
It is shown that the partial widths of the glueball production
in $J/\Psi$ radiative decays can be related to the vacuum to
glueball transition matrix elements calculated in this paper
\cite{lsl87,itll88,lli_future}. This is based on the approximation
that the charm quark is heavy so that $c$ and $\overline{c}$
annihilate at a point. The recent analysis of the $\pi\pi$ and
$K\overline{K}$ decays shows that a large component of spin zero
is observed in the $\theta(1720)$ region \cite{chen91}.
If this is confirmed, it would be a very good candidate for the
scalar glueball \cite{lli89}. The calculated scalar glueball
matrix element in eq. (\ref{eq:scalar}) predicts a branching
ratio of $5 \times 10^{-3}$ in $J/\Psi$ radiative decay
which is a only a few times larger than the
the experimental value from the observed decays in $\pi\pi$
and $K\overline{K}$ channels \cite{chen91}. The unobserved decay
modes will make the agreement better. The calculated pseudoscalar
matrix element gives a prediction of $4 \times 10^{-3}$ for
the branching ratio of the $0^{-}$ glueball candidate
$\eta(1440)$ in $J/\Psi$ radiative decay. This agrees quite
well with the observed B.R. of $\sim 3.5 \times 10^{-3}$ form the
decay modes of $K\overline{K}\pi, K\overline{K}^{\*}, a_0 (980)\pi$,
and $\rho^0\rho^0$ \cite{pdt92}.
On the other hand, the predicted branching
ratio for the tensor case, assuming that the $2^+$ component in
$\theta(1720)$ is the tensor glueball, is one to two orders of
magnitude larger than the experimental one. This could be due to
the fact that the tensor glueball matrix element is overestimated
as alluded to earlier. Or perhaps the tensor glueball
lies higher in mass (e.g. $>$ 2 GeV) leading to a smaller predicted
B.R.

To conclude, we have calculated the glueball to vacuum transition
matrix elements in quenched lattice QCD. Notwithstanding the fact that
our present study does not have high statistics and the Euclidean
time separation is limited, the variational method for the
hybrid correlation functions of the fuzzy and plaquette operators
does seem to yield consistent results for the glueball masses and
matrix elements. The matrix elements obtained for
the scalar and the pseudoscalar glueballs seem to agree reasonably
with those obtained phenomenologically and the prediction for
glueball production rates in $J/\Psi$ radiative decays are in line
with the experimental results based on the observed decay modes.
The tensor case could be an exception. It yields too large a
branching ratio for the tensor glueball in the $\theta$ region in
$J/\Psi$ radiative decay and is considerably larger than those
from the tensor dominance model and the QCD sum rule. We speculate
that it is due to the large finite size effect which is particularly
acute for the tensor due its large size \cite{fl92,fl92a}.
To verify this, one needs to conduct a high statistics calculation
on a larger lattice which should restore the rotational invariance
for the $E^{++}$ and $T_2^{++}$ and allow a larger separation in
t in order to have finite size effect under control.

\section{ ACKNOWLEDGEMENT} We thank W. Wilcox and C.M. Wu for helpful
discussions.  This work is partially supported by the DOE Grand
Challenge Award and DOE Grant No. DE-FG05-84ER40154.  Y.L. acknowledges
the support of the Center for Computational Sciences at the University
of Kentucky. K.I. is partially supported by a Grant-in-Aid for
general Scientific Research (03640256) and a Grant-in-Aid for
Scientific Research on Priority Area (04231101), the Ministry of
Education, Science, and Culture of Japan.

\begin{flushleft}
{\bf Figure Captions} \\

Fig. 1 \hspace{0.5cm} The making of fuzzy links {\it a la} Michael
     and Teper.

Fig. 2 \hspace{0.5cm} Shapes of different loops used in the
    variaitonal calculation.

Fig. 3 \hspace{0.5cm} Operators for the local field strength.
\end{flushleft}

\end{document}